\documentclass[useAMS,usenatbib]{mnras}

\usepackage{graphicx} 
\usepackage[latin1]{inputenc}
\usepackage{amsmath,amsfonts,amssymb}
\usepackage{color}
\usepackage{times}

\newcommand{\Msun}{\ensuremath{\textrm{M}_{\odot}}}

\newcommand{\kms}{km\hspace{0.25em}s$^{-1}$}

\newcommand{\MgI}{\mbox{Mg\hspace{0.25em}{\sc i}}}

\newcommand{\CaII}{\mbox{Ca\hspace{0.25em}{\sc ii}}}

\newcommand{\FeI}{\mbox{Fe\hspace{0.25em}{\sc i}}}
\newcommand{\FeII}{\mbox{Fe\hspace{0.25em}{\sc ii}}}
\newcommand{\FeIII}{\mbox{Fe\hspace{0.25em}{\sc iii}}}

\newcommand{\NiII}{\mbox{Ni\hspace{0.25em}{\sc ii}}}

\newcommand{\Fefsev}{$^{57}$Fe}
\newcommand{\Fefs}{$^{56}$Fe}
\newcommand{\Feff}{$^{54}$Fe}
\newcommand{\Cofs}{$^{56}$Co}
\newcommand{\Nifs}{$^{56}$Ni}
\newcommand{\Nife}{$^{58}$Ni}
\newcommand{\Mej}{M$_{\textrm{ej}}$}
\newcommand{\MCh}{M(Ch)}
\newcommand{\KE}{E$_{\rm K}$}

\newcommand{\eg}{e.g.,\ }
\newcommand{\ie}{i.e.,\ }

\def\gsim{\mathrel{\rlap{\lower 4pt \hbox{\hskip 1pt $\sim$}}\raise 1pt \hbox {$>$}}}
\def\lsim{\mathrel{\rlap{\lower 4pt \hbox{\hskip 1pt $\sim$}}\raise 1pt \hbox {$<$}}}
\def\gtaprx {\lower .1ex\hbox{\rlap{\raise .6ex\hbox{\hskip .3ex
	{\ifmmode{\scriptscriptstyle >}\else
		{$\scriptscriptstyle >$}\fi}}}
	\kern -.4ex{\ifmmode{\scriptscriptstyle \sim}\else
		{$\scriptscriptstyle\sim$}\fi}}}
\def\ltaprx {\lower .1ex\hbox{\rlap{\raise .6ex\hbox{\hskip .3ex
	{\ifmmode{\scriptscriptstyle <}\else
		{$\scriptscriptstyle <$}\fi}}}
	\kern -.4ex{\ifmmode{\scriptscriptstyle \sim}\else
		{$\scriptscriptstyle\sim$}\fi}}}

\begin{document}

\title[Clumping in SN\,2014J]
  {The intermediate nebular phase of SN\,2014J: onset of clumping as the 
  source of recombination}
\author[P. A.~Mazzali et al.]{P. A. Mazzali$^{1,2}$
\thanks{E-mail: P.Mazzali@ljmu.ac.uk}, 
I. Bikmaev$^{3,4}$, R. Sunyaev$^{2}$, C. Ashall$^5$, S. Prentice$^6$, 
M. Tanaka$^7$, 
\newauthor E. Irtuganov$^{3,4}$, S. Melnikov$^{3,4}$, R. Zhuchkov$^{3,4}$\\
\\
  $^{1}$Astrophysics Research Institute, Liverpool John Moores University, IC2, Liverpool Science Park, 146 Brownlow Hill, Liverpool L3 5RF, UK\\
  $^{2}$Max-Planck-Institut f\"ur Astrophysik, Karl-Schwarzschild Str. 1, 
  	D-85748 Garching, Germany\\
  $^{3}$Department of Astronomy and Satellite Geodesy, Kazan Federal 
  	University, Kremlevskaya Str., 18, Kazan, 420008, Russia\\
  $^{4}$Academy of Sciences of the Republic of Tatarstan, Bauman Str., 20, Kazan, 
  420111, Russia \\
  $^{5}$Florida State University, Tallahassee, FL, USA \\
  $^{6}$Trinity College Dublin, Ireland \\
  $^{7}$Tohoku Univ., Sendai, Japan
}

\date{Accepted ... Received ...; in original form ...}
\pubyear{2019}
\volume{}
\pagerange{}

\maketitle

\begin{abstract}  
At the age of about 1 year, the spectra of most Type Ia supernovae are dominated
by strong forbidden nebular emission lines of \FeII\ and \FeIII. Later
observations (at about 2 years) of the nearby SN\,2011fe showed an unexpected
shift of ionization to \FeI\ and \FeII. Spectra of the very nearby SN Ia 2014J
at an intermediate phase (1 -- 1.5 years) that are presented here show a
progressive decline of \FeIII\ emission, while \FeI\ is not yet strong.
The decrease in ionization can be explained if the degree of clumping in the ejecta 
increases significantly at $\sim 1.5$ years, at least in the Fe-dominated zone.  Models suggest that clumps remain coherent after about one year, behaving like shrapnel. The high density in the clumps, combined with the decreasing heating rate, would then cause recombination. These data may witness the phase of transition from relatively smooth ejecta to the very clumpy morphology that is typical of SN remnants. The origin of the increased clumping may be the development of local magnetic fields.  
\end{abstract}


\section{Introduction}
\label{sec:introduction}

Many issues remain unclear about Type Ia supernovae (SNe\,Ia), although there
is agreement that they are the thermonuclear explosion of carbon-oxygen (CO)
white dwarfs (WD), and that most of them produce a large amount of radioactive \Nifs. A relation between \Nifs\ mass, luminosity and light curve shape was
discovered \citep{phillips93} and explained in terms of the behaviour of the
opacity \citep{pintoeastmann2000,mazzali2001a,zorro}. This relation has been
applied so that SNe\,Ia can be used as standardizable candles, leading to the
discovery of the accelerated expansion of the Universe
\citep{riess98,perlmutter99}.  

Many different progenitor scenarios have been proposed for SNe\,Ia. A short list
includes: i) the single-degenerate (SD) scenario, where a WD accretes matter from a
non-electron degenerate donor companion in a binary system
\citep[\eg][]{whelaniben73,nom84w7}; ii) the double-degenerate scenario (DD), which
consists of two WDs in a binary system that eventually merge
\citep{webbink84,ibentutukov84}; iii) the ``core-degenerate" scenario, where a WD is
engulfed in a common envelope by the outer layers of a binary companion when this
expands to become an asymptotic giant branch (AGB) star until the WD finally merges with the degenerate core of the AGB star \citep[\eg][]{livioriess2003,kashisoker2011}, and iv) the explosion triggered by the head-on collision of two WDs, either in a dense environment such as a cluster \citep{Rosswog09} in a triple system where the third member is a non-electron-degenerate star, which focuses the orbits of the two WDs \citep{kushnir13}. For a recent review of all scenarios see
\citet{liviomazz2018rev}.

Additionally, within each progenitor system a variety of different explosion mechanisms have been
proposed to occur.  When a WD approaches the Chandrasekhar mass (\MCh) through
accretion from a companion star, compressional heating triggers a thermonuclear
runaway in the centre. The flame can propagate at both sub- and super-sonic speeds,
leading to different nucleosynthesis and energy yields. In this scenario the
companion can be a degenerate or a non-degenerate star
\citep[\eg][]{Iwamoto99,Piersanti03}.  In sub-\MCh\ explosions an accreted surface He layer can ignite and cause a shock wave on the surface of the WD. If the shock propagates into the WD it can produce a central detonation. Once again this mechanism can occur in both the SD and DD scenarios, through accretion from a He star or another WD with a He layer \citep{livne95,Shen14}. 

Although supporters of the different scenarios often claim that their favourite
model can explain all SNe\,Ia, it is not unlikely that different, physically plausible channels to a SN\,Ia actually co-exist. While models of the early-time properties of SNe\,Ia do not always lead to conclusive results about the specific progenitor/explosion channel \citep[\eg][]{roepke12}, clearer evidence for the existence of multiple channels comes from nebular spectroscopy
\citep[\eg][]{zorro,mazzali2015_11fe_neb,mazzali_03hv,Mazzali_07on,mazzali_91bg_neb}.

In the late, nebular phase, when the inner parts of the ejecta are visible,
distinguishing among different explosion scenarios based on their expected
imprint is somewhat easier. This is because the central density, and hence the 
total mass of the exploding WD directly affects the nucleosynthesis and the
density distribution in the inner-most regions of the ejecta, and different models tend to exhibit the largest differences in the inner layers. 



In addition to outstanding questions about the nature of the progenitor system and the
mode of explosion, the link between the SNe\,Ia we observe as luminous transient
sources and the SN remnants (SNR) they produce also offers interesting puzzles. SN\,Ia remnants are thought to be quite spherical and homogeneous, which should reflect the properties of the explosion and testify to the lack of significant circumstellar medium (CSM) to shape the remnant at later stages \citep{Patnaude17}.  While this is in line with expectations for progenitor evolution \citep[although in some scenarios, such as WD collisions and mergers, the ejecta may not be expected to be spherically symmetric, \eg][]{bulla2016_DD}, those SNRs that are identified as most likely being the result of a SN\,Ia explosion are characterised by significant clumping. \citet{ferrand19} argue that clumps can be developed by Rayleigh-Taylor instabilities, over a timescale of $\sim 100$ yr. However, the presence of clumping in the first few years after the explosion is not suggested by either the explosion models or by observations in the early phases.

The period in a SN intermediate between light curve peak and the SNR is the so-called 
nebular phase. This phase is entered when the SN ejecta start to become optically
thin, some six months after the SN explosion, and continues as long as collisional
excitation and ionization determine the state of the gas and before impact on any CSM
lights up the nebula in a reverse shock and shapes the SNR if the CSM is not
spherically symmetric. Because SNe\,Ia have only been observed in external galaxies in the modern era, rarely has it been possible to follow them in the nebular phase much beyond one year, as they quickly become too faint to be observed spectroscopically. 

On the other hand, most well-studied SN remnants are in the Milky Way or the
Magellanic Clouds, where they are close enough to be observed. Typical SNRs have
ages of 1000 years or more. Therefore, the transition from SN to SNR is
practically unexplored. The occurrence in recent years of two SNe\,Ia in nearby
galaxies offers an opportunity to explore more advanced stages of the nebular
phase than were previously known. SN\,2011fe, a normal SN\,Ia
\citep{nugent2011}, was monitored in the early nebular phase (up to one year
after explosion). Its spectrum showed strong emission lines of both [\FeII] and [\FeIII]. The spectra were shown to be consistent with a Chanderasekhar-mass
explosion \citep{mazzali2015_11fe_neb}. At more advanced epochs the light curve showed the onset of different radioactivity \citep[in particular
\Fefsev,][]{tauben_15_11fe}, as expected from nucleosynthesis calculations.
Surprisingly, however, SN\,2011fe displayed a shift in ionization, with spectra
taken 593 and 981 days after the explosion still showing [\FeII] lines but not
[\FeIII] lines, which seemed to have been replaced by [\FeI] lines
\citep{graham2015}. A noisy optical spectrum obtained 481 days after explosion 
\citep{zhang2016} appears to show the two strong [\FeII] and [\FeIII] 
emissions near 4800 and 5200\,\AA, but also an emission near 4500\,\AA, which 
\citet{zhang2016} identify as [\FeII] but which, based on its wavelength, may also be consistent with the [\FeI] emission that is seen at very late times in SN\,2011fe. Although it would be unusual for three different ionization stages to be strong at the same time, this is the only evidence of how the transition from [\FeIII]- to [\FeI]-dominated spectra in SN\,2011fe may have occurred. While some degree of cooling could be expected as the SN ejecta age and expand, the degree of recombination that has been observed in SN\,2011fe was not predicted by models, 
which favour a freezout in ionization as the density decreases \citep{frajerk2015}. 

Another very nearby, recent SN\,Ia is SN\,2014J, which exploded in M82 and was
the closest SN\,Ia since 1972E. Several papers have focussed on the early-time 
properties and the unusual extinction of SN\,2014J
\citep[\eg][]{foley2014,Amanullah14,ashall14J}.  
Nebular spectra of SN\,2014J have been presented in a few papers. 
\citet{srivastav2016} presented two spectra taken at epochs $\approx 288$ and 370 days after explosion. They noticed the weakness of the [\FeIII]-dominated emission relative to other SNe\,Ia and interpreted this as a signature of clumping in the ejecta following \citet{mazzali2001b}. 
Two papers have dealt specifically with late-time, near-infrared (NIR) spectra.
\citet{Diamond18} conclude that SN\,2014J was consistent with a \MCh\
progenitor, which exploded as a delayed detonation with an off-centre ignition.
\citet{Dhawan18} claim to have observed a [\NiII] 1.939\,$\micron$ feature,
consistent with high density central burning, which again would be consistent with a \MCh\ progenitor. Here we present late-time optical spectra of SN\,2014J 
that span epochs from 10 to 14 months after explosion. The spectra do show a 
progressive decrease of [\FeIII] emission, although [\FeI] is not emerging yet, 
and thus cover the initial part of the ionization shift. Modelling the spectra we 
obtain an approximate description of the properties of the ejecta and of their 
changes during the period monitored. These changes lead to recombination in the 
ejecta and the progressive fading of [\FeIII] emission.

\section{Spectroscopic observations, data subtraction and calibration}
\label{sec:data}

\begin{table*}
\caption{Log of observations for SN2014J with RTT-150}
\label{table:obslog}
\centering
\begin{tabular}{lccc}
Date &	JD & Days after $B$ maximum & Total Exposure per night (s) \\
\hline
2014 Nov 11 &	2456973	& 283 & 	4500 \\
2014 Dec  4 &	2546996 & 306 & 	5400 \\
2014 Dec 14 &	2457006	& 316 &		2700 \\
\hline
Average     &   2456993	& 303 &	Total exposure 12600 sec  \\
\hline
\hline
2015 Jan 21 &   2457044	& 354 &		9000  \\
\hline
\hline
2015 Mar 24 &	2457106	& 416 & 	9000 \\
2015 Mar 25 &	2457107	& 417 &		5400 \\
\hline
Average	    &   2457106 & 416 &	Total exposure 14400 sec \\
\hline
\end{tabular}
\end{table*}

Observations of SN\,2014J were performed on the 1.5-m Russian-Turkish
telescope RTT-150 \citep{aslan2001} using time allocated to Kazan Federal
University. The low-resolution spectroscopic mode of the TFOSC instrument was
used. A log of observations is given in Table \ref{table:obslog}. A prism in 
combination with a 100 micron (1.8 arcsec) entrance slit and a 2048 x 2048 
pixels nitrogen-cooled CCD provide a spectral resolution of 15\,\AA\ in the
wavelength range 4000--9000\,\AA. The spectra were reduced using a modified 
version of the DECH software package\footnote{www.gazinur.com}. The wavelength
scale was calibrated using the comparison spectra of an Fe-Ar lamp. To calibrate 
the fluxes in the spectra, the spectrophotometric standard stars BD+75d325 and 
Feige 34 were observed. 

The wavelength calibration accuracy is $\approx 0.1$\,\AA\ ($\approx 5$\,\kms).
An additional uncertainty of $\approx 50$\,\kms\ was caused by the motion of the
stellar image perpendicular to the spectrograph entrance slit during the exposure. 
To increase the signal-to-noise ratio in the spectra of SN\,2014J at a late nebular 
phase, the spectra taken on November 11, December 4 and 14, 2014, were averaged 
(the average spectrum corresponds to day 303 after maximum brightness), as were the 
spectra taken on March 24 and 25, 2015, (the average spectrum corresponds to day 
416 after maximum brightness).

The observed spectrum of SN\,2014J during the late nebular phase is contaminated by
underlying host-galaxy light. In order to correct for this, a spectrum of the host at
the position of the SN is required at a time when the SN light makes a negligible
contribution to the total observed flux. A single 800\,s spectrum of M82 at the
coordinates of SN\,2014J was therefore obtained on 2019 Jan 14.03 (MJD 58497.03)
using the SPectrograph for the Rapid Acquisition of Transients
\citep[SPRAT;][]{piascik2014}, installed on the 2\,m Liverpool Telescope
\citep[LT;][]{steele2004}.  SPRAT is a low resolution, high throughput spectrograph which covers a wavelength range of 4000 -- 8000\,\AA\ and reaches a binned resolution of R=350 at the centre of the spectrum for a slit width of 1.8 arcsec. The late epoch of this exposure ensured that the SN had faded from view. The data were initially reduced through the standard SPRAT pipeline, which applies corrections for bias, dark, and flat-field. The object spectrum was then extracted, and calibrated in wavelength and flux using a custom {\sc pyraf} pipeline. The extracted spectrum was left unbinned, giving a resolution of R=634 ($\approx 9$\,\AA) at 6000\,\AA. A final flux correction, derived from the standard star used to define the sensitivity function, was applied to the output spectrum to correct for telluric absorption. 

The spectra of SN 2014J were calibrated in flux with respect to the Hubble Space
Telescope (HST) photometry published by \citet{yang2018}.  Because only 3
photometric HST bands (F475W, F606W, and F775W) are available, we assumed that
the spectra had the correct shape after reduction and proceeded to calibrate
them in flux by correcting them using a constant to ensure that they matched
the observed photometry.  

\begin{figure*} 
\includegraphics[width=139mm]{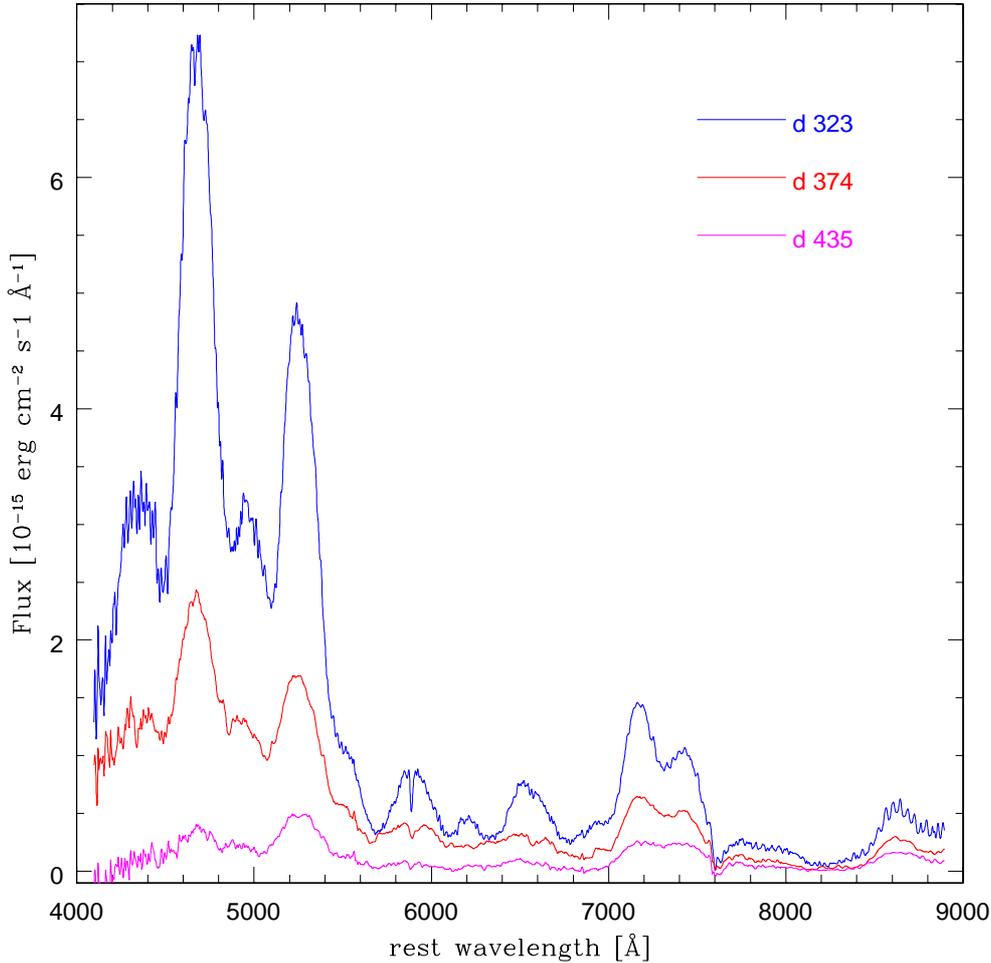}
\caption{The three nebular spectra of SN\,2014J used in this work, in their original flux after correcting for reddening. The spectral sequence highlights the decline of the emission near 4700\,\AA\ relative to other emission features.}
\label{fig:allnebspe}
\end{figure*}

The spectra were then corrected for extinction and reduced to restframe. For the Milky Way we used $E(B-V)=0.05$\,mag, $R_V=3.1$, and for the host galaxy $E(B-V)= 1.2$\,mag, $R_V=1.38$ \citep{ashall14J}. This is consistent with the extinction values found in \citep{foley2014, Amanullah14}. The spectral sequence, after correction for extinction and removal of the host galaxy spectrum, is shown in Figure \ref{fig:allnebspe}.

As Fig. \ref{fig:allnebspe} shows, over the period sampled by our data the spectrum of SN\,2014J showed a significant change in degree of ionization. The first two epoch
were dominated by blends of [\FeIII] and [\FeII] emission lines, as is typical of SNe\,Ia  in the classical nebular phase (up to about one year).  At the epoch of the third spectrum however, although the main emission features remained the same as at earlier epochs, the ratio of the two strongest Fe emission changed. While at earlier epochs the [\FeIII]-dominated emission near 4700\,\AA\ was significantly stronger than the [\FeII]-dominated one near 5200\,\AA, at day 435 the ratio of these two features is reversed, with the  [\FeII]-dominated one now the dominant emission. This may be seen as the beginning of a transition similar to that which occurred in SN\,2011fe, which showed strong [\FeI] emission at an epoch of $\sim 3$ years after explosion. [\FeI] emission is not observed in the available spectra of SN\,2014J, but it may be expected that it could appear, or even become dominant, at later times. Thus motivated, we undetook an analysis of the spectra of SN\,2014J using our SN nebular code to explore what could lead to the observed evolution of the spectrum.

\section{The SN Nebular Code}
\label{sec:code}

We computed synthetic spectra using our non-local thermodynamic (NLTE) nebular emission code for supernovae. The code is built based on the assumptions set out by \citet{axelrod80}. The gas in the SN nebula is assumed to be heated by collisions with the high-energy particles generated in the thermalization process of the gamma-rays and positrons emitted in the decay chain \Nifs\ $\rightarrow$ \Cofs\ $\rightarrow$ \Fefs, and it cools via the emission of (mostly) forbidden lines. Some strong permitted transitions are also considered.

The deposition of gamma-rays and positrons is computed using a Montecarlo method,
as outlined in \citet{cappellaro97} and \citet{mazzali2001a}. Constant opacities are used of $\kappa_{\gamma}=0.027$\,cm$^2$g$^{-1}$ and
$\kappa_{e^+}=7$\,cm$^2$g$^{-1}$. The large difference in their respective opacities means that at very late times positrons make the dominant contribution to energy deposition, although they carry only $\approx 3.6$\% of the total emitted radioactive energy.

After computing the energy deposition, the ionization and the NLTE thermal balance in the nebula are solved following the prescriptions outlined by, \eg \citet{ruizlaplucy92}. Ionization is assumed to be provided exclusively by impact with the high-energy particles produced by the deposition of gamma-rays and positrons. Photoionization is assumed to be negligible \citep{kozmafransson98}. Ionization balance is achieved by equating the rate of impact ionization with the recombination rate for each ion. Within each ion, level populations are computed solving the rate equations for each level under the assumption of thermal balance, \ie\ equating the non-thermal heating rate and the rate of cooling via line emission. The assumption is made that the nebula is optically thin, and radiation transport in not performed. The emissivity in the various lines is used to compute the emerging spectrum. 

The code can be run as a simple one-zone code, in which case the line profile is a parabola with characteristic width determined by the speed of expansion of the outer boundary of the nebula (which is input to the code). Alternatively, a stratified version can use a one-dimensional explosion model (\ie\ a radially dependent density/abundance profile), in which case energy deposition, ionization balance, level populations and emissivity are computed for each shell. Each shell then contributes its own emissivity. Line profiles are built by summing truncated parabolae bounded by each shell's inner and outer boundary velocity. In both cases, the effects of line blending are automatically taken into account in the computation of the emitted spectrum.

Our code has been used for a number of SNe Ia \citep[\eg][]{mazzali_91bg_neb,mazzali_03hv} and Ib/c \citep[\eg][]{mazzali06ajneb}, and it can yield a description of the inner layers of the SN ejecta. 
When used in combination with a description of the outer layers, which can be obtained through a study of the early spectra, the mapping of the inner ejecta that results from nebular studies can be used to obtain a full description of the density and abundance distribution in a SN \citep[\eg][]{mazzali2015_11fe_neb,ashall_86G,ashall11iv}.

One aspect our code attempts to treat is clumping of the ejecta. The simple assumption is made that all ejecta reside in clumps, which occupy a fraction of the total volume, as defined by a filling factor. The density in the clumps is then increased following the decrease in volume, and the composition of the clumps is set to be the same as that of the radial ejecta shell they are in, \ie\ all clumps are homogeneously and microscopically mixed. The clumps are still assumed to be in thermal equilibrium. These are of course major simplifications, but in the inner layers of a SN\,Ia only few elements are present, so it is not unreasonable. Strong clumping has been shown to be very useful in order to reproduce the low ionization seen in late-time spectra of SNe\,Ic \citep[\eg]{mazzali2001b}, where it may be thought of as a proxy for the non-spherically symmetric distribution of the ejecta. Clumping is used later in this paper to improve our synthetic spectra, especially in the final epoch.

\section{Spectral Models and the evolving properties of SN\,2014J}
\label{sec:models}

We started our analysis by deriving some basic properties of the emitting nebula using the one-zone version of our code. The flux progressively
decreases with time, as is expected for a spectrum that is powered by the decay of
\Cofs, but the width of the emission lines decreases only marginally,
suggesting that over the time monitored the nebular emitting region remains
roughly constant. This confirms the results of \citet{mazzali98}.

From simple 1-zone fits to the spectra we estimate a typical expansion velocity of the emitting nebula of $8500 \pm 500$\,\kms. This places SN\,2014J among relatively luminous SNe\,Ia, according to the scheme of \citet{zorro}. For normal SNe\,Ia this emission line width corresponds to decline rates of $\Delta(m)_{15} \sim (1.1 - 1.3)$\,mags. 
This is reasonably consistent with the observed value for SN\,2014J, which had an extinction-corrected decline rate $\Delta(m)_{15} = 1.08$\,mags. SNe\,Ia with this range of decline rates typically have a \Nifs\ mass of $\sim (0.4 - 0.6)$\,\Msun\ \citep{zorro}. 

A much more accurate description of the properties of the young SN nebula can be obtained using the stratified version of the code in combination with an explosion model. As an example, using this approach we have been able to show that the peculiar behaviour of the nebular specta of some underluminous SNe\,Ia, namely the unexpected {\em increase} in ionization at late times, can be explained if the innermost ejecta contain little mass, and in particular no stable Fe, such as predicted by sub-Chandrasekhar mass models as a consequence of the low central density of the progenitor white dwarfs \citep{mazzali_03hv,mazzali_91bg_neb}. The low density suppresses recombination, and the absence of stable Fe reduces the cooling rate, leading to the observed effects. In the case of SN\,2014J the behaviour is the opposite. 

The question then arises of which explosion model we should select. A study of the properties of SN\,2014J based on the early spectral evolution and light curve was performed by \citet{ashall14J}. They fitted a series of spectra and concluded that SN\,2014J was consistent with a normal, Chandrasekhar-mass SN\,Ia, with kinetic energy \KE$\,\sim 1.2\; 10^{51}$\,erg. The classical, albeit artificially fast deflagration model W7 \citep{nom84w7} proved perfectly adequate to reproduce the early spectra, requiring only minor modifications to the abundances. That work could only probe the outer ejecta, down to a velocity of $\sim 8500$\,\kms. 

Here, therefore, we follow \citet{ashall14J} and adopt the density structure of W7 to sample the lowest-velocity regions of the ejecta. As nebular-epoch spectroscopy sees in principle ``through" the entire ejecta, which at that point have low optical depth at most frequencies, being able to model three different epochs allows us on the one hand to optimize our results as regards the abundance distribution as a function of radius, and on the other to identify any behaviour that is not simply the result of the evolution in time of the SN nebula, as determined by the progressive decrease in density. It should be noted here that, because of its artificial nature, W7 resembles the density and abundance distribution of delayed detonation models \citep[\eg][]{Iwamoto99} in the inner layers, such that it would be practically  impossible to distinguish among them based on nebular spectra only. The main differences in the density distribution are found in the outer layers, reflecting the different energetics of the different models, while differences in composition are mainly at intermediate velocities, where again burning depends on energy. Therefore, based on the result of \citet{ashall14J}, who favour a relatively low-energy explosion, we use the W7 density model in our simulations (but retain the freedom to modify the abundance distribution somewhat). We also note that, as a matter of fact, delayed detonation mdoels are also artificial, as the transition from subsonic to supersonic burning is typically introduced arbitrarily.

We computed nebular spectra to match the observed spectra of SN\,2014J using our NLTE nebular code. In SNe\,Ia most of the cooling occurs via forbidden lines of Fe, but some permitted and semi-forbidden transitions of lighter elements are also active coolants (\eg the \CaII\ IR triplet and H\&K lines, \CaII]\,7291,7324\,\AA, \MgI]\,4570\,\AA). A [\NiII]\,7380\,\AA emission line is also expected if stable Ni is produced in the explosion, and is weakly observed in the spectra of SN\,2014J.

\begin{figure*} 
\includegraphics[width=139mm]{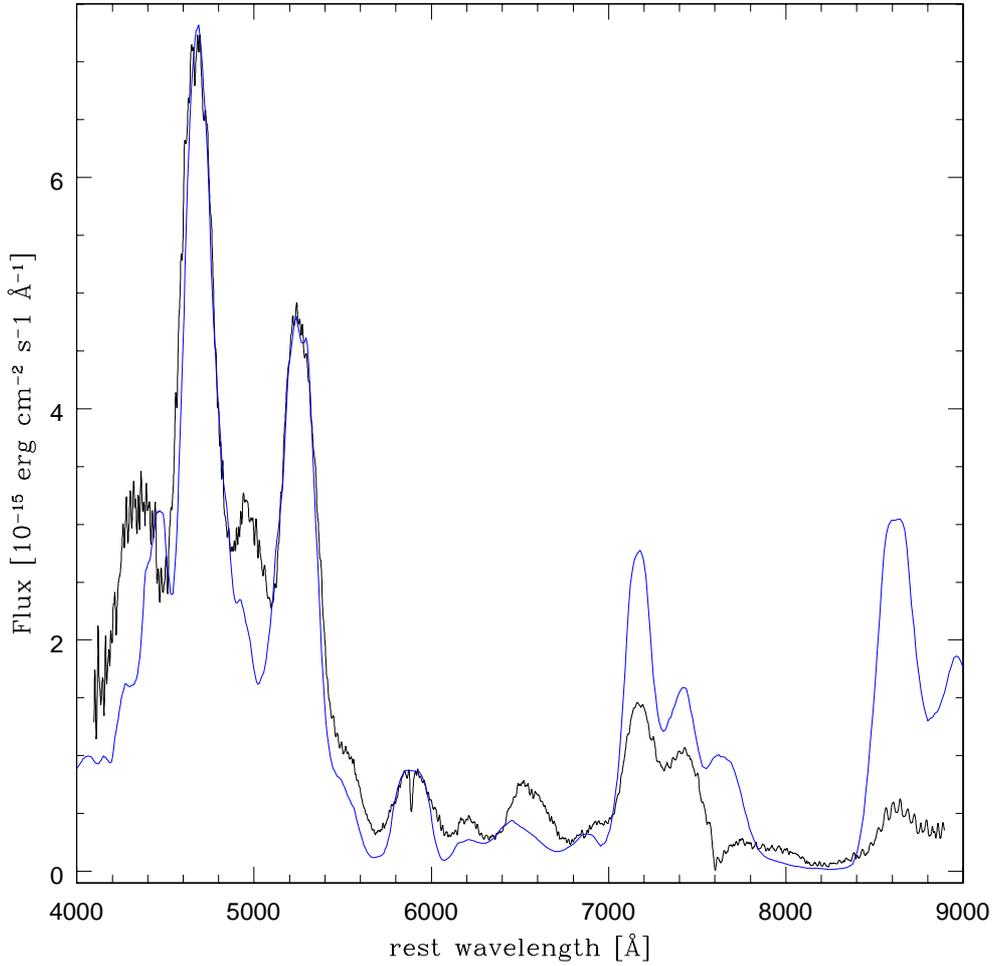}
\caption{The synthetic nebular spectrum based on the 1D explosion model with stratified density and composition and optimised to match the features of SN\,2014J at 323 rest-frame days after explosion (blue), compared to the observed SN spectrum after de-reddening and subtraction of the host spectrum (black). }
\label{fig:d323spec_strat}
\end{figure*}

Our modelling strategy reflects our standard approach for nebular spectral
series: we first reproduced the earliest spectrum, and then used the same
parameters (density and composition as functions of radius/velocity) derived
from that fit to model the later epochs. The assumption behind this approach is
that the ejecta are transparent at all nebular epochs. If this is true, all
spectra should be reproduced by changing the epoch only, which allows for
expansion and more radioactive decay. This has usually been successful for
time-series of spectra, of both SNe\,Ia and Ib/c
\citep[\eg][]{mazzali07_02ap_neb,mazzali2015_11fe_neb}. Therefore, we start by
discussing the fit of the spectrum taken 303 rest-frame days after $B$-band maximum. 
A rise time of 20 days \citep{ashall14J} has to be included in our models, so that the density can be properly rescaled. The three spectra have therefore epochs of 323, 374, and 435 rest-frame days after explosion, respectively. For all our models we assumed  a distance modulus $\mu = 27.86$\,mag, corresponding to a distance of 3.77\,Mpc, following \citet{ashall14J}.

The spectrum at 323 days (Fig. \ref{fig:d323spec_strat}) looks just like any other nebular spectrum of a normal SN\,Ia. The two strongest emission features (after correction for reddening) are two blends of Fe lines in the blue. The feature near 4800\,\AA\ is dominated by [\FeIII] (strongest lines at 4658, 4701, 4734\,\AA), while the one near 5200\,\AA\ is composed mainly of [\FeII] lines (strongest ones at 5159, 5262, 5336\,\AA), but they both contain lines of both ions (\eg [\FeII] lines at 4815, 4890\,\AA, [\FeIII] lines at 5270, 5412\,\AA). This means that it is not possible to tune the ratio of the strength of these two emission features, which is controlled by the ionization ratio and the populations of the upper levels of the emitting transitions and therefore depends on both temperature and density. Of these two factors, the former has a secondary effect though, as level populations are not in LTE. In this case, as was noted several times previously, the ionization degree is determined by the competition between heating and cooling, which depends on density and therefore - indirectly - also on composition. At the typical densities of SN\,Ia ejecta at an age of about 1 year, a gas composed exclusively of \Nifs\ tends to be hot and highly ionised, as a significant part of it contributes to heating (via \Cofs\ decay). Therefore, \FeIII\ is likely to prevail. However, this situation is actually quite rare in SNe\,Ia, and whenever it is observed it is in SNe that are peculiar spectroscopically, tend to be subluminous, and are best explained as sub-\MCh\ events \citep[\eg][]{mazzali_91bg_neb,mazzali_03hv}.  These SNe have in common a low density in the inner layers, which disfavours recombination, and so do the scenarios to which they have been associated.

\begin{figure} 
\includegraphics[width=84mm]{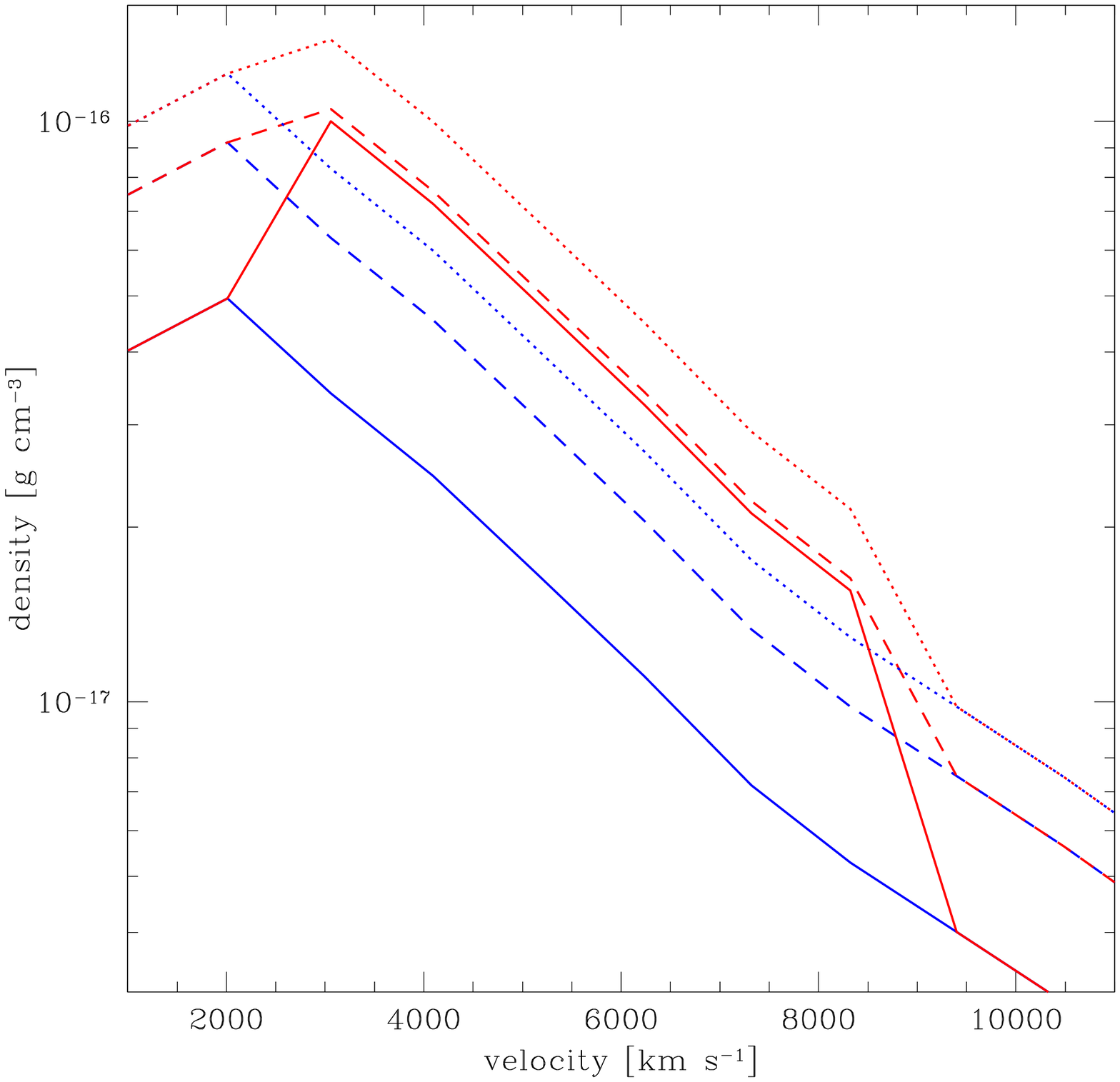}
\caption{Density as a function of velocity in the Ni-rich ejecta for different epochs. The blue lines are the density of W7 at the times of the models, 323, 374 and 435 days from top to bottm, respectively, while the red lines show the density in the clumps. }
\label{fig:density}
\end{figure}

\begin{figure} 
\includegraphics[width=84mm]{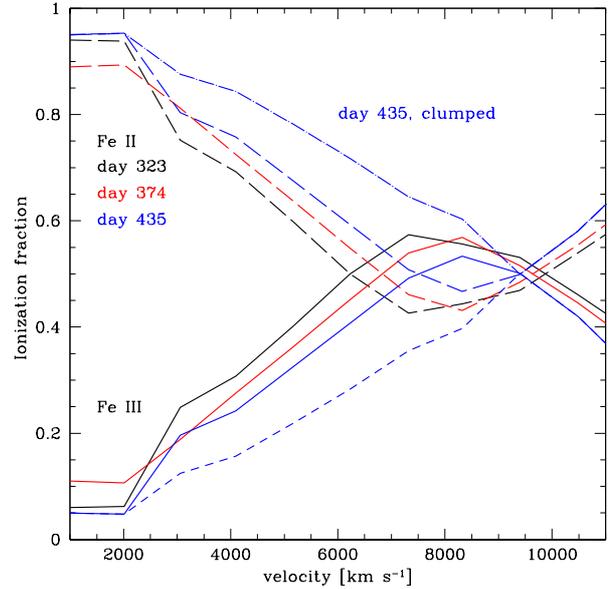}
\caption{The ionization of Fe as a function of velocity in the ejecta for different epochs as computed by our model. The strongly clumped model for day 435 results in much lower ionization. The plot is limited to regions where energy deposition from gamma-rays and positrons is significant and leads to emission in Fe lines.}
\label{fig:Feion}
\end{figure}

\begin{figure*} 
\includegraphics[width=139mm]{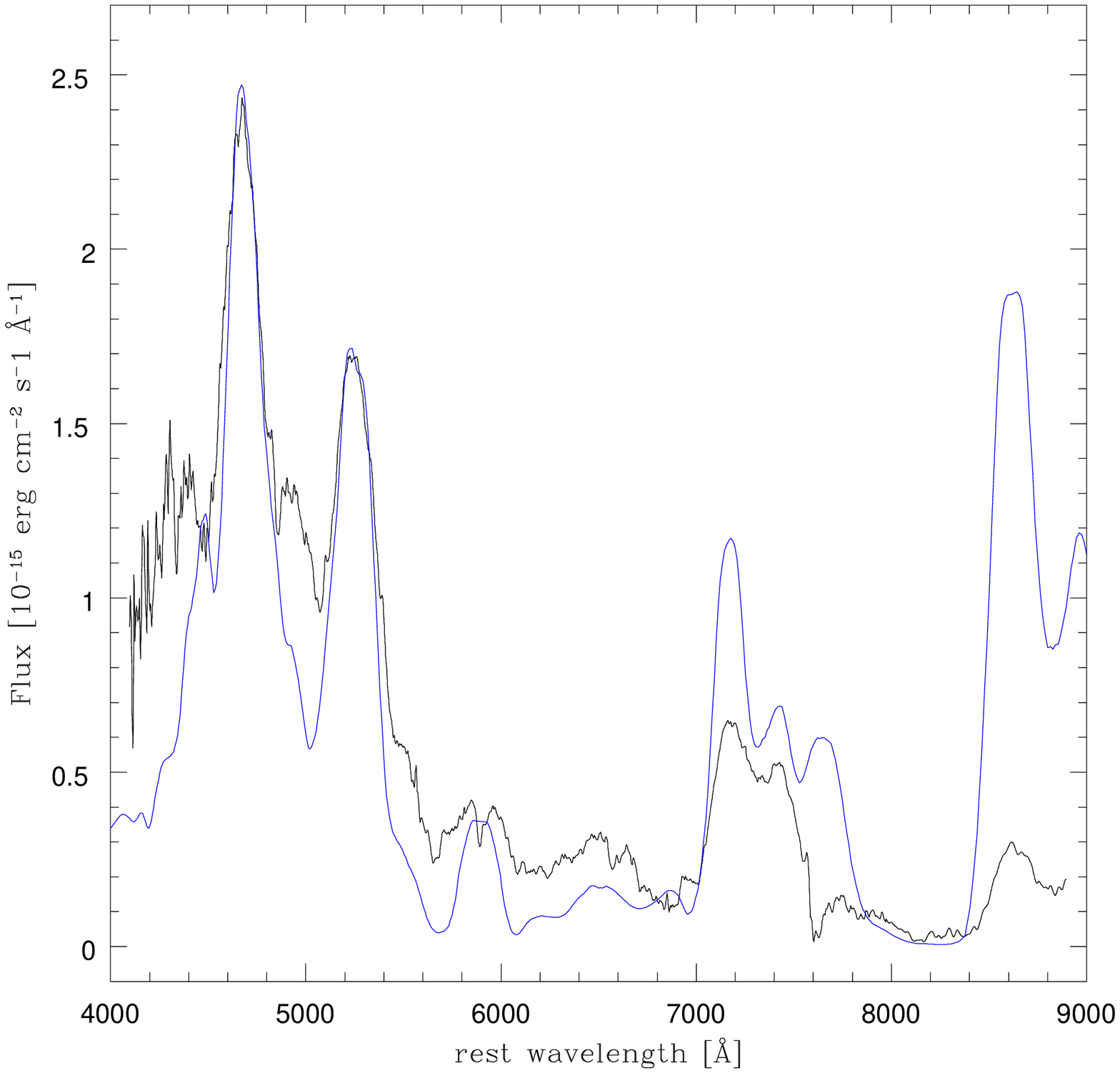}
\caption{The synthetic nebular spectrum based on the 1D explosion model with stratified density and composition and optimised to match the features of SN\,2014J at 374 rest-frame days after explosion (blue), compared to the observed SN spectrum after de-reddening and subtraction of the host spectrum (black). }
\label{fig:d374spec_strat}
\end{figure*}

The most natural way to cause the gas to achieve a lower degree of ionization appears to be adding stable iron. This effectively cools the ejecta, as stable Fe can be heated by radioactive decay products and then contributes to cooling via line emission, but is not itself the product of a process that produces any heating. Additionally, stable Fe cools via the same emission lines that Fe deriving from \Cofs\ decay emits, thereby not causing unwanted spectral features. Introducing stable Fe in the Ni-rich ejecta indeed works well for all normal SNe\,Ia, and this is the case for SN\,2014J, too. Calculations of explosive nucleosynthesis in near \MCh\ SNe\,Ia suggest that neutron-rich, stable isotopes of Fe and Ni (typically \Feff\ and \Nife) are synthesized at the highest densities, in the centre of the WD \citep[\eg][]{Iwamoto99}, but they are present with significant abundance (especially \Feff) also in the \Nifs-dominated region. Sub-\MCh\ WDs do not reach the high densities necessary to synthesize stable Fe and Ni, and therefore the presence or absence of stable Fe is an important element to discriminate between these two classes of models. The presence of stable Ni can be tested at late times via the emission of [\NiII] 7380, as radioactive \Nifs\ has almost completely decayed away at one year. On the other hand, emission from stable Fe just adds to the lines emitted by Fe produced by \Cofs\ decay, and as we cannot spectroscopically distinguish between different isotopes a direct measurement of the quantity of Fe synthesised as stable is not possible. A solution must be found whereby the balance between heating and cooling provides the correct strength and ratio of [\FeII] and [\FeIII] lines, thereby matching both the \Nifs\ production and the observed ionization within the volume comprised by the velocity of the Fe emission lines. 
A good balance between emission line intensity and the ratio of the iron ions that radiate was found in the case of SN\,2014J for a \Nifs\ mass of $\approx 0.46$\,\Msun, located mostly at velocities between 3000 and 9000\,\kms, and a stable Fe mass of $\approx 0.24$\,\Msun, located mostly at velocities between 1000 and 7000\,\kms. This is actually in line with typical explosion models of a Chandrasekhar-mass CO WD undergoing a delayed detonation \citep[\eg][]{Iwamoto99}. 

The results above are also in line with the expectations based on \citet{zorro}.  The mass of \Nifs\ is consistent with the decline rate of SN\,2014J. The mass of stable Fe is also in line with other normal SNe\,Ia. It is required to fill the density structure of W7 as well as to keep the ionization sufficiently low. The fact that a good ionization balance is reached at the density of a \MCh\ model with a typical stable Fe content indeed simultaneously supports both the W7 density distribution and a \MCh\ explosion as the most likely scenario for SN\,2014J. 
At 323 days \FeII\ and \FeIII\ are present in similar amounts, with a ratio that depends on radius. The runs of density and Fe ionization as a function of depth are shown for this and the other models discussed in this paper in Figures \ref{fig:density} and \ref{fig:Feion}, respectively. When the innermost parts of the density profile of W7 are filled with \Nifs\ and stable Fe a reasonable ionization balance is reached, but best results are achieved if it is further assumed that the \Nifs-rich region is also clumped. Very moderate clumping is required (volume filling factors of $\sim 0.60$), similar to other normal SNe\,Ia, in regions where \Nifs\ dominates the abundance. This leads to a slightly lower ionization of Fe. Confining all \Nifs\ to lower velocities (higher densities) to favour recombination would actually change the width of the emission lines and yield unsuitable synthetic spectra \citep{mazzali2019}. 

The stable Ni mass, on the other hand, is quite small, $\approx
0.005$\,\Msun, based on the intensity of the [\NiII] 7380 line. This result is in contrast with typical \MCh\ explosion models \citep[\eg][]{Iwamoto99}, but is not unusual for normal SNe\,Ia, as we discuss below. Stable Ni has to be located mostly below 2000\,\kms\ in order to avoid an excessively strong [\NiII] 7380 line. It is worth noting here that the [\NiII] line is not observed to have any shift with respect to its expected rest-frame position. Considering that SN\,2014J is a marginal member of the low velocity gradient \citep[LVG;][]{benetti2005} group of SNe\,Ia \citep[$\dot v = -58.4 \pm7.3$\kms d$^{-1}$;][]{galbany2016}, this is not in contradiction with a possible 3D picture of most SNe\,Ia as off-centre delayed detonation explosions \citep{maeda2010}.

\begin{figure*} 
\includegraphics[width=139mm]{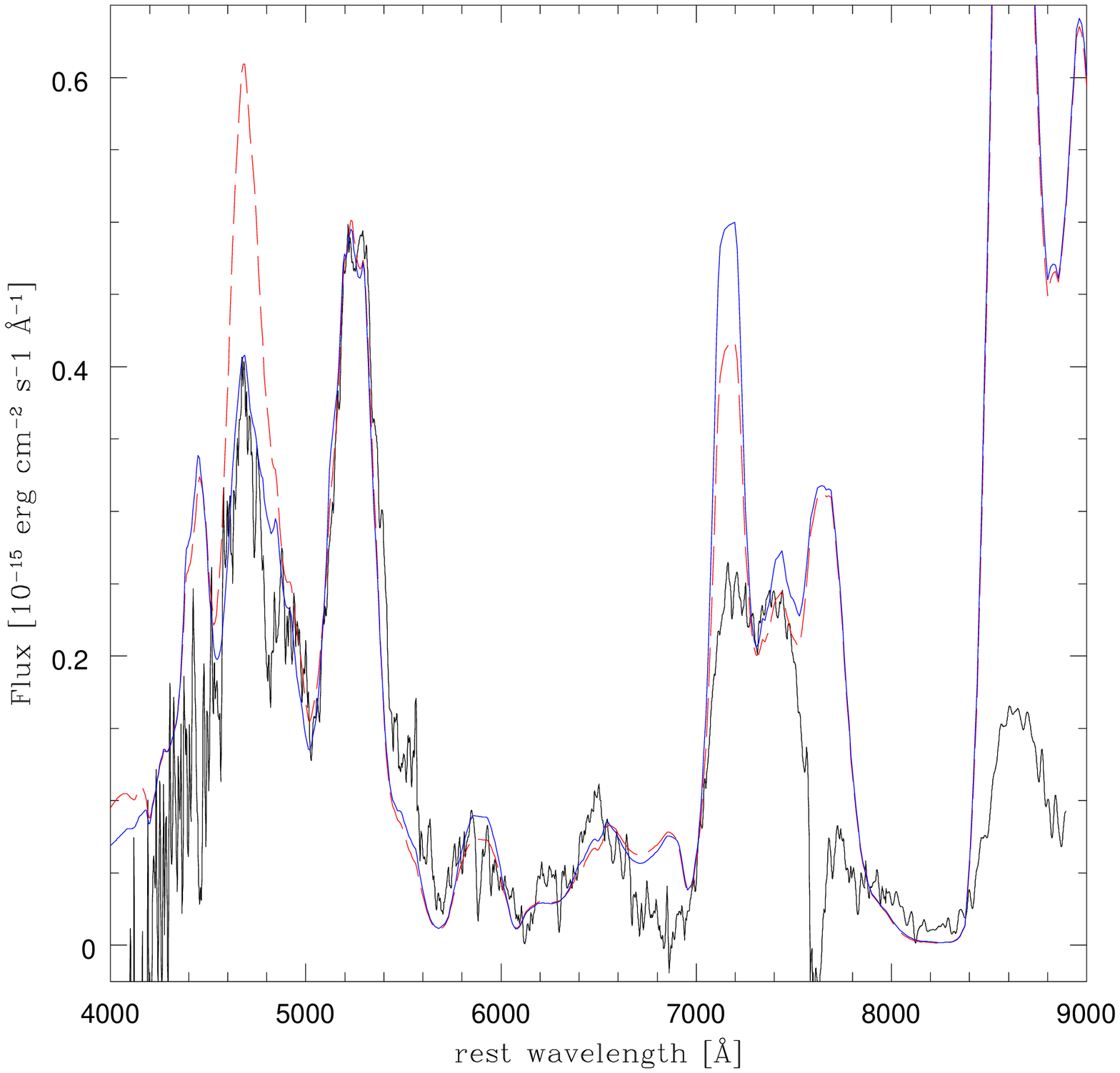}
\caption{Synthetic nebular spectra based on the 1D explosion model with stratified density and composition, compared to the observed SN spectrum of SN\,2014J at 435 rest-frame days after explosion after de-reddening and subtraction of the host spectrum (black).  The dashed red line show the model with stratified density, composition and clumping as in previous epochs. The full blue line shows the 1D model with stratified density and composition as in previous epochs but enhanced clumping to suppress \FeIII.}
\label{fig:d435spec_strat}
\end{figure*}

Regions in the ejecta lying outside a velocity of $\approx 7500$\,\kms\ were
assumed to be characterised by the abundance distribution determined from the
models of the early-time spectra in \citet{ashall14J}. These regions are
dominated by intermediate-mass elements (IME), in particular silicon and
sulphur. They are separated from \Nifs\ and therefore are not subject to much heating and do not radiate significantly. The ionization is therefore lower in those regions, as shown in Fig.\ref{fig:Feion}.

Having determined a likely solution for the d323 spectrum, we now move to the next epoch, d374. The expectation is that simply modifying the epoch should yield a good fit to the observed spectrum. This is indeed the case: Fig. \ref{fig:d374spec_strat} shows the observed spectrum of SN\,2014J after dereddening and subtraction of the host galaxy spectrum together with the synthetic spectrum obtained using the same density and abundance distribution as on day 323, as well as the same clumping. The success of this test indicates that the model of the density and abundance distribution that we adopted is reasonably correct. There are a large number of SN\,Ia\ nebular spectra at epochs near 1 year, and the spectra of SN\,2014J are similar to those of other normal SNe\,Ia.

Finally, we model the last spectrum of our series. This has an epoch of 435 days
after explosion, and thus it samples an advanced epoch that had not been monitored in other SNe\,Ia, except for the spectrum of SN\,2011fe at 481 days after explosion published by \citet{zhang2016} which is discussed above.  It is characterised by the onset of the drop of the [\FeIII] emission, although no strong [\FeI] emission is yet visible. 

If we use the same model as for the previous two epochs and just change the reference time we obtain the synthetic spectrum shown as a dashed line in Fig.
\ref{fig:d435spec_strat}. Although the synthetic spectrum does show a drop in
[\FeIII] emission, this is much less than what is actually observed. If no changes occur, the predicted degree of ionization in the model at this epoch is naturally lower than at earlier epochs because of the decreasing efficiency of gamma-ray deposition in the lower density nebula (see Fig. \ref{fig:Feion}), but still high enough for significant [\FeIII] emission to be present, which is in conflict with observations. A similar behaviour was noted at even much later epochs in the models of \citet{frajerk2015}.

\begin{figure*} 
\includegraphics[scale=0.6,angle=90]{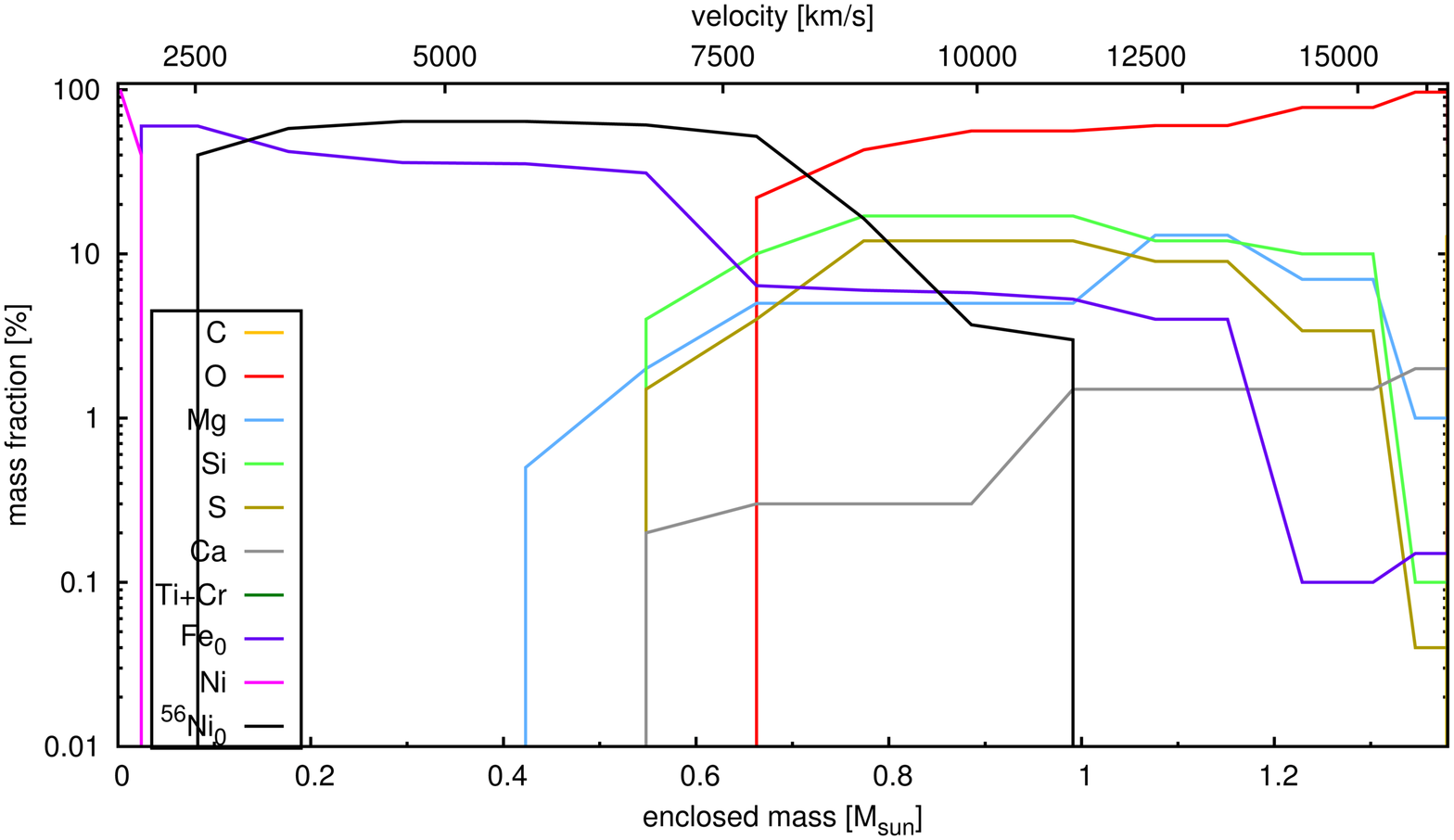}
\caption{The abundance distribution in the ejecta of SN\,2014J as determined combining the early-time results of \citet{ashall14J} above $v = 7500$\,\kms\ and the nebular result from the present paper at lower velocities.}
\label{fig:Abund}
\end{figure*}

If we want to keep the radial density and abundance distribution that was successful
at earlier phases and at the same time ensure a lower degree of ionzation, and thus a
better match to the observed spectrum, the only change that we can make at this later
epoch is to increase clumping. \citet{frajerk2015} also note in passing that clumping may improve the match between their synthetic spectra and the very late spectra of SN\,2011fe (they address much later epochs than we do here, when [\FeIII] emission has completely disappeared). 

Our approach to clumping affects in particular the regions that are rich in \Nifs, namely the inner ejecta. We do not separate different elements in the clumps, so that heating is largely unaffected (at these late epochs most energy deposition is due to positrons, which deposit locally as they typically have a short mean free path, and Fe coexists even at the microscopic level with \Nifs\ and hence with the sites where positrons are created). Cooling takes place via the usual emission lines, but the higher densities in the clumps enhance recombination there, such that equilibrium is achieved at a lower ionization regime than in an unclumped or less clumped medium.

If we increase clumping significantly at the latest epoch a shift in ionization
occurs and the observations can be reproduced. This can be achieved using a
volume filling factor $ff = 0.30$ in the \Nifs-rich ejecta at $t=435$ days. The corresponding synthetic spectrum is shown in Fig. \ref{fig:d435spec_strat} as a thick blue line. Increasing clumping implicitly means increasing the density in the regions of the ejecta that actively emit. The density as a function of velocity is shown in Fig. \ref{fig:density} for the three epochs, both without clumping and with the clumping that yields best fits to the Fe emission. Interestingly, the density at d435 when enhanced clumping is used is similar to the density of the clumped model at d374, so that the differences in ionization degree between the two epochs are due primarily to the lower rate of ionization. The enhanced clumping at d435 leads to significant changes in Fe ionization with respect to the model with the same degree of clumpping as at previous epochs, as can be seen in Fig. \ref{fig:Feion}. For example, at $t=435$ days, the ratio \FeII/\FeIII\ reaches values of 2-3 in the shells that mostly contribute to emission in the highly clumped model, while in the model with moderate clumping it is more like 1-2. With a lower ionization, [\FeIII] emission is significantly reduced, and the synthetic spectrum reproduces the observations quite well.  Although we had to increase clumping to suppress [\FeIII] emission, the degree of clumpiness we used is not yet extreme, possibly indicating that the ejecta are undergoing a transition to a more clumpy structure.  In the blue the spectrum has a lower signal-to-noise ratio. However, the difference between the clumped and unclumped models is clearly visible, and the more clumped model matches the data much better.

The abundances and clumping properties of the regions outside those where \Nifs\ dominates cannot be assessed using this method, as those regions are inefficiently heated by radioactive decay products at late epochs and are therefore largely passive. The abundance distribution in the ejecta of SN\,2014J obtained by combining the early-time results of \citet{ashall14J} and the nebular results obtained here is shown in Fig. \ref{fig:Abund}. It is consistent with the results of several other normal SNe\,Ia, \eg\ 2011fe \citep{mazzali2015_11fe_neb} or 2004eo \citep{mazzali_04eo}, although the production of Si and S appears to be somewhat suppressed. These results come from the analysis of the early time data in \citet{ashall14J}, and we do not discuss them again here.

\section{Aside: Could extreme clumping be used at all epochs?}

\begin{figure*} \includegraphics[width=139mm]{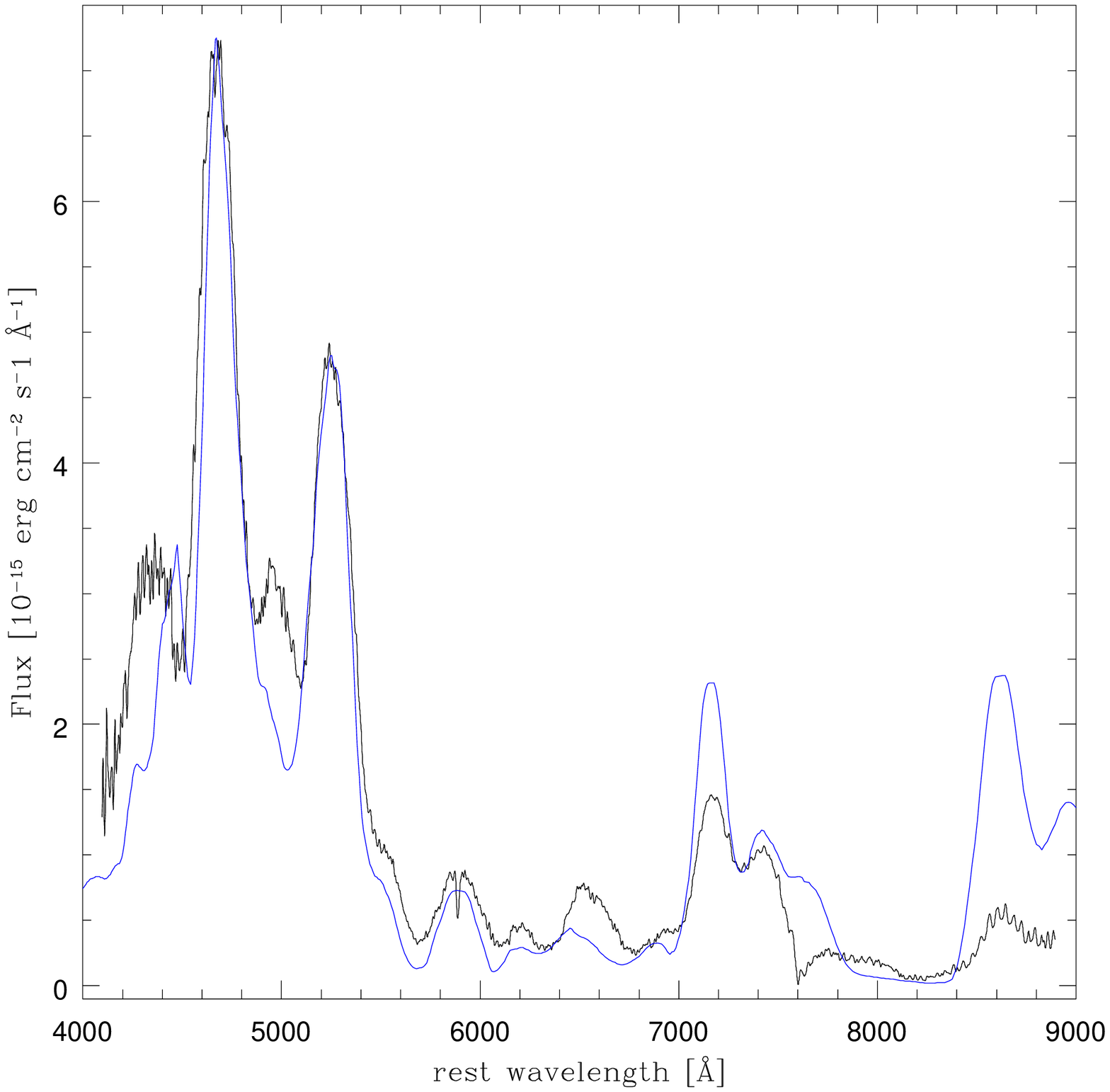}
\caption{The synthetic nebular spectrum based on the 1D explosion model with ejected mass 1.1\,\Msun, stratified density and composition, compared to the observed SN spectrum  at 323 rest-frame days after explosion after de-reddening and subtraction of the host spectrum (black). }
\label{fig:subch323}
\end{figure*}

\begin{figure*} \includegraphics[width=139mm]{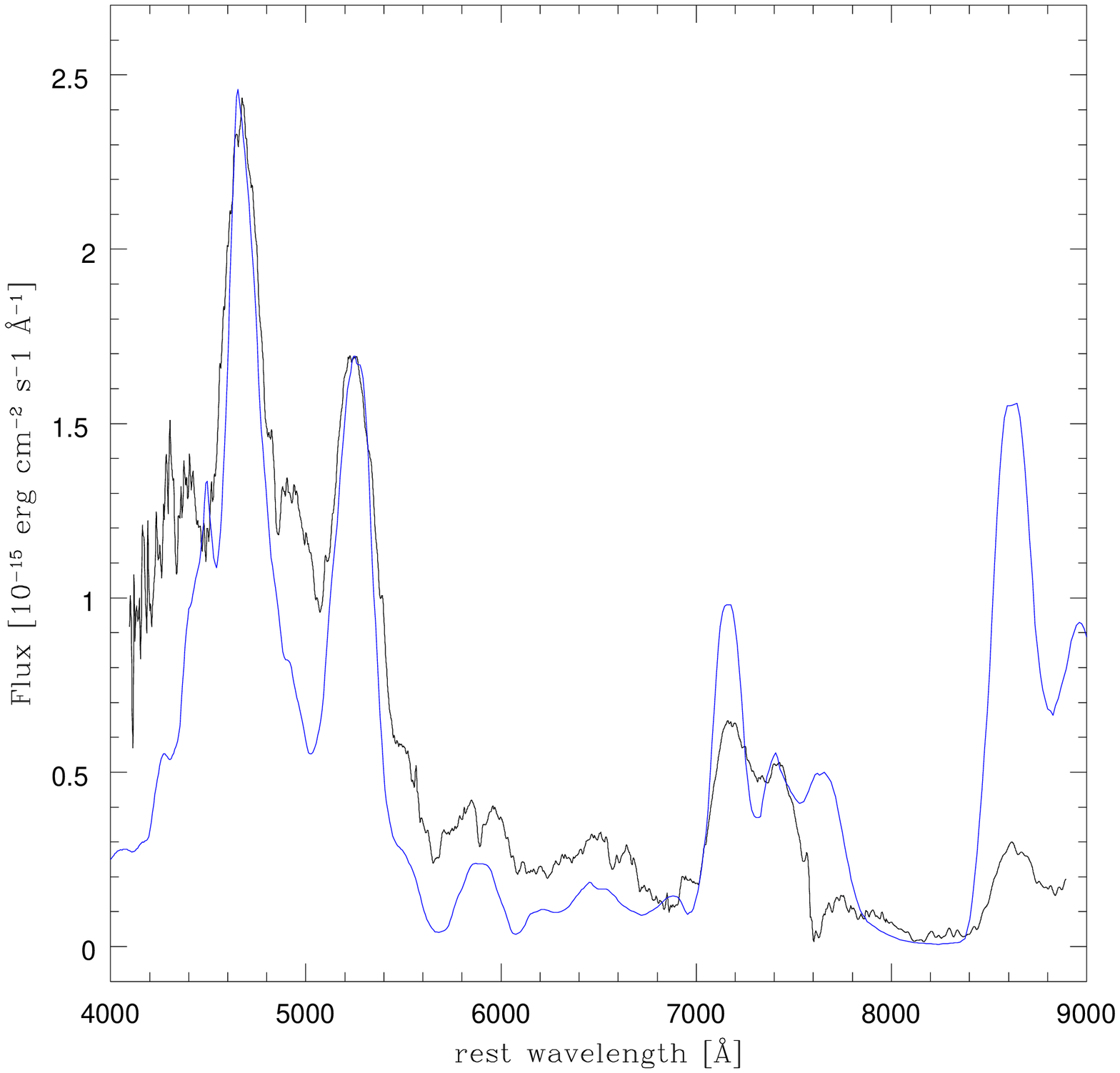}
\caption{The synthetic nebular spectrum based on the 1D explosion model with ejected mass 1.1\,\Msun, stratified density and composition, compared to the observed SN spectrum  at 374 rest-frame days after explosion after de-reddening and subtraction of the host spectrum (black). }
\label{fig:subch374}
\end{figure*}

\begin{figure*} \includegraphics[width=139mm]{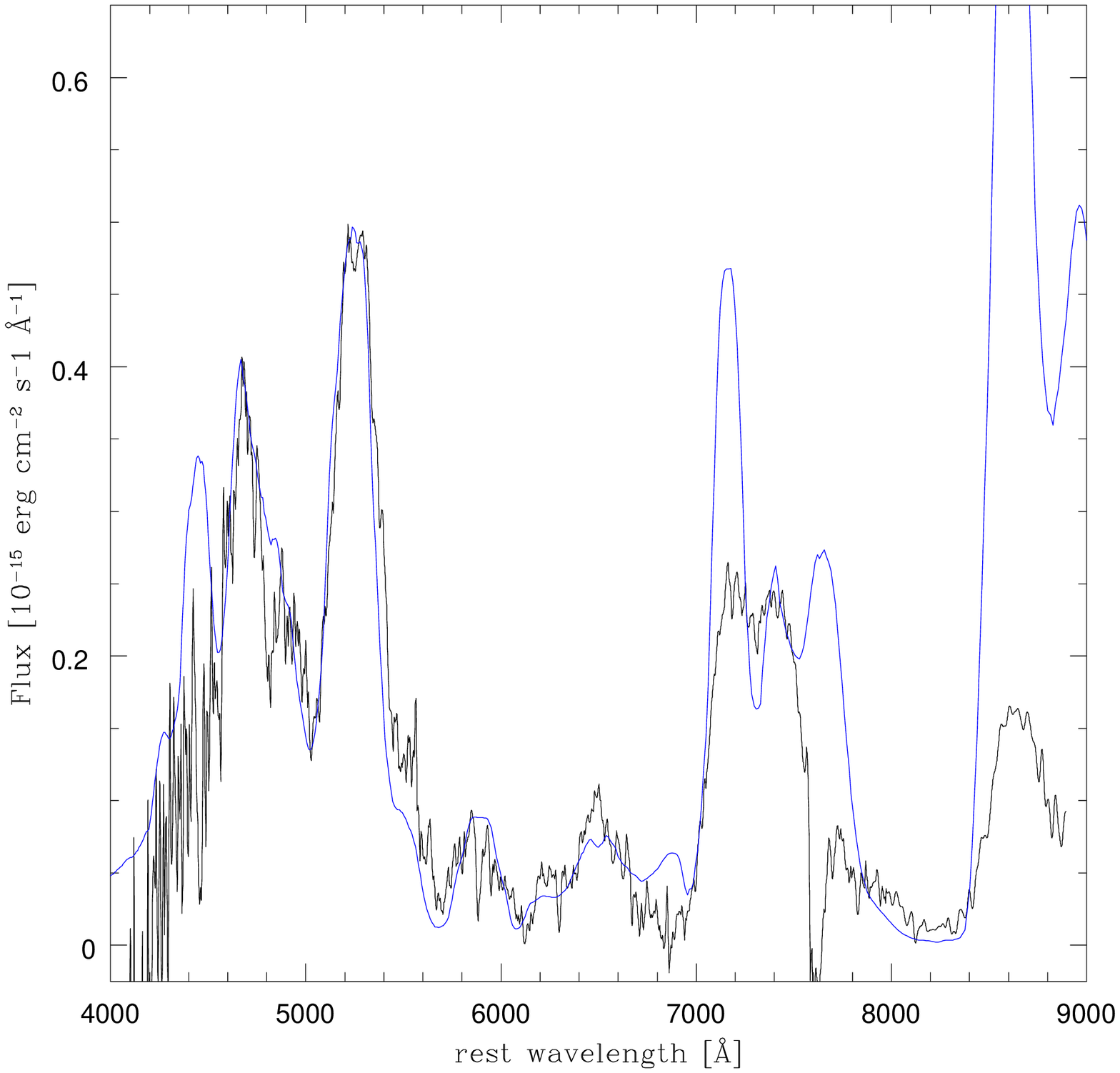}
\caption{The synthetic nebular spectrum based on the 1D explosion model with ejected mass 1.1\,\Msun, stratified density and composition, compared to the observed SN spectrum  at 435 rest-frame days after explosion after de-reddening and subtraction of the host spectrum (black). }
\label{fig:subch435}
\end{figure*}

Given that we invoked clumping to suppress [\FeIII] emission at very late times, one question that naturally arises is could clumping be more extreme than previously
estimated also at earlier epochs during the nebular phase. More extreme clumping may
be seen as a way to achieve a lower ionization, and therefore the correct balance of
[\FeIII] and [\FeII] emission, without having to resort to stable Fe to provide
additional cooling. This seems to be the situation for example for SNe\,Ib/c, where
stable Fe is not supposed to be synthesized during the explosion and yet strong [\FeIII] emission is seen at no point during the nebular phase. Filling factors 
of the order of 0.1 are required to suppress [\FeIII] emission in SNe\,Ib/c \citep[\eg][]{mazzali07_02ap_neb}, much more extreme than in the case of SN\,2014J. However, SNe\,Ib/c are also thought to be highly aspherical
\citep[\eg][]{maeda06,ashall19}, so clumping may also just mimic the effect 
of a non-smoothly spherical symmetric distribution of \Nifs\ when modelling is performed in one dimension \citep[\eg][]{mazzali2015_11fe_neb}.

In the case of SN\,2014J, we asked the question: if we increase clumping, eliminate
stable Fe and thus just reduce the mass in the ejecta, can a consistent reproduction
of the data be obtained? The three steps above are necessary in order to decrease
ionization while avoiding cooling via unwanted emission lines of other elements that
may be introduced when trying to conserve mass.  What we describe here amounts to
constructing a model with the characteristics of a sub-\MCh\ explosion, a scenario
that is favoured by some for the bulk of SNe\,Ia because searches for both surviving
companions and the remains of accreted hydrogen, both of which are important features
of the single degenerate scenario, have thus far failed to yield positive results. It
shouod be noted, however, that possible ways to avoid those issues exist for both of
these expected observations \citep[\eg][]{distefano2011,distefano2012,justham2011}.

We attempted to produce a reasonable synthetic nebular spectrum at
$t=323$ days for an ejected mass \Mej$\sim 1.1$\,\Msun.
This value of \Mej\ is in line with many sub-\MCh\ models 
\citep[\eg][]{Sim2010_subCh}. 
A resonable nebular spectrum could be obtained for a \Nifs\ mass of
$\sim 0.44$\,\Msun\ if we used quite an extreme clumping, \ie\ a filling factor $ff = 0.28$. The reduced ejected mass comes at the expense of the amount of material in the inner ejecta. In particular, the mass of stable Fe is only 0.03\,\Msun, while that of stable Ni remains very small, $\approx 0.002$\,\Msun, because of the weakness of the [\NiII]\,7380 line.

The synthetic spectrum thus obtained is shown in Fig.\ref{fig:subch323}.  The quality of the fit is similar to that of the Chandrasekhar-mass model, which may be seen as a positive aspect for this solution. The only slight difference is that the Fe emission lines are somewhat narrower. 
A reduced core mass with practically unchanged outer layers is indeed a feature of
sub-Chandrasekhar-mass models \citep[see][]{mazzali_03hv}. However, it also would probably imply a shorter diffusion time of photons at early times, and result in a narrower light curve at peak, as has indeed been observed for peculiar, subluminous SNe\,Ia \citep[see][]{mazzali_03hv}.  It also must be noted that clumping is not currently a feature of sub-\MCh\ models either, let alone strong clumping.

However, this model is not stable, both in clumping and in \Nifs\ mass. The strong clumping adopted for the model at $t = 323$ days favours gamma-ray deposition, especially in the inner layers, such that if clumping remains constant in time then the later epochs are predicted to have higher emission flux than is actually observed. In order to fit the spectra the \Nifs\ mass must be reduced as a function of time, which is unlikely to be an acceptable solution. At $t=374$ days the \Nifs\ mass has to decrease to 0.40\,\Msun\ for a reasonable fit, which is however not very good. A better fit can be achieved if clumping is slightly reduced, to $ff = 0.32$ (see Fig.\ref{fig:subch374}). The stable Fe mass is now larger, 0.06\,\Msun, while the stable Ni mass is still 0.002\,\Msun. At both of these epochs, in addition to the problems mentioned above, the removal of mass makes it difficult to produce broad enough Fe emission features, even though the inner regions (up to $v \approx 7000$\,\kms\ are now dominated by \Nifs. This is because of the missing Fe emission and the reduced mass. 

At $t = 435$ days, however, in order to obtain a good fit, the \Nifs\ mass must increase again somewhat (now 0.42\,\Msun, at the expense of stable Fe, now 0.04\,\Msun) but clumping must again increase significantly, now to $ff = 0.18$ (see Fig.\ref{fig:subch435}). So, irrespective of the mass adopted, strong clumping is needed at the late epoch. 

Is the sub-\MCh\ scenario a viable solution?
The somewhat poorer fits to the data on the one hand, together with the unphysical requirement of a time-dependent \Nifs\ on the other, make us favour the Chandrasekhar-mass model with increasing clumping at the latest epoch, as fewer time-dependent modifications are necessary in this case. Additionally, a reduced mass is likely to lead to a more rapidly evolving light curve near peak. The kinetic energy of the sub-\MCh\ model was not greatly reduced, and mass was removed only from the deepest, lowest-velocity layers, both in order to test the nebular spectra and to keep the density/abundance structure that was successfully used to fit the early-time spectra \citep{ashall14J}.

\section{Discussion}
\label{sec:disc}

Our results indicate that the increase of clumping in the ejecta of SN\,2014J at epochs past one year after explosion may have been responsible for the decrease in ionization which in turn is reflected in the evolving spectral properties, with a drop in [\FeIII] emission. 

The presence of significant clumping at the time of the latest of our three nebular
spectra of SN\,2014J suggests that the ejecta evolve morphologically, making a
transition from a smoother regime which applies earlier on to a more clumpy state. In
SN\,2014J this transition appears to start about one year after the explosion. It may
continue into later times, leading to even more extreme clumping, but unfortunately
later data are not available for SN\,2014J. Evidence from SN\,2011fe is that
eventually [\FeIII] lines disappear and are replaced by [\FeI] lines, which suggests that clumping may be active in that SN, too, as suggested by \citet{frajerk2015}.
Evidence for increased clumping in SN\,2014J comes from the spectrum at 435 days.
Previous work on nebular spectra of SNe\,Ia had been limited to spectra with epochs
barely exceeding one year \citep[\eg][]{zorro}, and such a shift in ionization  was not seen in any of the normal SNe\,Ia that were analysed in that work. In the future it will be important to make every effort to follow SNe into the nebular phase, and well after one year whenever possible, in order to establish whether the behaviour of SN\,2014J (and SN\,2011fe) is actually common to the bulk of (at least normal) SNe\,Ia.

Our description of the clumpy properties of SN\,2014J is a very simple one. It is not
easy to extract such information from the spectrum. Our one-dimensional modelling has
clear limitations, but if one resorted to 3D modelling the number of free parameters
(clump size, density, composition, distribution as well as any underlying deviation
from spherical symmetry in the explosion model) would be so large that finding a
solution may be quite hard, and multiple solutions may be found. As we mentioned
above, current SN\,Ia explosion models do not predict the development of clumps, but
this is a feature that future models may have. For the SN\,IIP-peculiar 1987A
\citet{jerk87A} used a multi-dimensional model to inform their simulations. No
available SN\,Ia model predicts either clumpy ejecta or the development of clumps with time, so for the time being our approach appears at least to bring out the nature of the physical process at play, although not its details. 

\begin{table*}
\caption{Properties of one-zone models}
\label{table:onezone}
\centering
\begin{tabular}{ccccccc}
Epoch 	& M(\Nifs) & M(Fe) & M(stable Ni) & M$_{tot}$ & velocity & ff \\
(days) & [\Msun] & [\Msun]& [\Msun] & [\Msun]& [km s $^{-1}$] &\mbox{~} \\
\hline
323	&  0.66	     & 0.04  &  0.01	 & 0.76    &  7700 & 0.50 \\
374	&  0.67	     & 0.03  &  0.02	 & 0.77    &  9000 & 0.50 \\
435	&  0.73	     & 0.03  &  0.02	 & 0.83    &  8000 & 0.50 \\
\hline
\end{tabular}
\end{table*}

An interesting scenario is suggested looking at Fig. \ref{fig:density}: the model with enhanced clumping at d435 has very similar densities as the moderately clumped model at d374. This suggests that clumps may form, or be present in the ejecta, about one year after the explosion, and that they do not expand, behaving more like shrapnel than small gas clouds. This may indeed be consistent with the observed properties of SNRs.

A possible cause for the onset of clumping could be the development of magnetic fields
caused perhaps by the relative motion of charged particles such as the ions and
electrons in the ejecta. The young SN nebula is significantly ionized, reaching
electron densities up to $\sim 10^6$\,cm$^{-3}$ in the inner ejecta at 435 days in the model with enhanced clumping, so this is a possibility, especially in the presence of velocity offsets between different species. The actual details are likely to be quite complex, and are certainly well beyond the possibilities of our spectrum synthesis approach and the scope of this paper. Magneto-hydrodynamic calculations would be required, and those are encouraged. 

The small mass of stable Ni is in contrast with typical Chandrasekhar-mass model predictions. Such models \citep[\eg][]{Iwamoto99} typically foresee a ratio of stable Ni to stable Fe of $\sim 1$. In all Chandrasekhar-mass models we have computed \citep[\eg][]{zorro,mazzali2015_11fe_neb} this ratio is actually much smaller than 1.  The small Ni mass stems from the weakness of the only observed Ni line in the nebular spectrum, [\NiII]\,7380, which we can fit with a stable Ni mass of $\approx 0.005$\,\Msun.
A low Ni/Fe ratio was also recently obtained by \citet{floers2020}, who claim that it suggests that the bulk of SNe\,Ia are sub-\MCh\ explosions. However, sub-\MCh\ explosions also produce very little stable Fe directly: most stable Fe is just the result of the decay of \Nifs. \citet{floers2020} use one-zone models to determine the final ratio of Fe (including both stable Fe directly synthesized and Fe produced by \Nifs\ decay). We computed one-zone models for the three nebular spectra of SN\,2014J presented here in order to test the results. 
We could obtain reasonable fits for the three spectra using the parameters listed in Table \ref{table:onezone}. The fits were driven by the effort to match the strength and width of the strongest emission lines, which are complex blends, as we remarked above. One-zone models are very useful to determine the actual extent of the effectively emitting zone, as they take line blending into account. Just like \citet{floers2020}, we find that the stable Ni mass is small, and in addition we also do not require much stable Fe. However, in a one-zone models the real conditions in the ejecta are not well reproduced. In particular, the density is constant within the emitting sphere, which is far from reality. In our one-zone models we adopted a filling factor of 0.5 as a proxy for the density distribution, 
otherwise we could not achieve a balance between the emitting flux, the width of the lines, and the ionization balance. 
In our 1D models we find a small mass of stable Ni because the single visible Ni line, [\NiII]\,7380, is weak, but this is a typical feature of all SNe\,Ia. However, the mass of directly made stable Fe is large, and the total mass of stable Fe-group elements produced is actually consistent with the prediction of Chandrasekhar-mass explosion models.  We also find that stable Ni tends to be at lower velocities than directly made stable Fe as well as \Nifs, which is again consistent with Chandrasekhar-mass models. One possibility is that differences between the central density estimated in models and their real value may lead to a different nucleosysnthesis than predicted.

\section{Conclusions}
\label{sec:concl}

We have modelled three nebular spectra of SN\,2014J. We find that a good match to the observed spectra can be obtained if SN\,2014J was a Chandrasekhar-mass explosion. The SN produced $\sim 0.45$\,\Msun\ of \Nifs. It also produced some 0.3\,\Msun\ of stable Fe, and a small amount ($\sim 0.005$\,\Msun) of stable Ni. The mass of \Nifs\ that we obtain is in line with expectations for a SN\,Ia with the luminosity of SN\,2014J. The mass of stable Ni is small compared to the predictions of Chandrasekhar-mass explosion models, but the mass of stable Fe is larger than in those models, such that the combined mass of stable Fe-group material produced directly is in line with model predictions. A low mass of stable Ni is found in all SN\,Ia whose nebular spectra we have modelled, including those compatible with the Chandrasekhar-mass. This suggests that SN\,2014J was a Chandrasekhar-mass explosion, and that the detailed nucleosynthetic predictions of some explosion models are not correct.

In order to match the sharp decrease in emission intensity of the \FeIII-dominated
feature near 4800\,\AA\ at the latest epoch, 435 days after explosion, we had to
increase clumping at that epoch. Although our approach to clumping is very simplistic,
namely we assume that all material is within the clumps and completely mixed according
to the radial distribution of abundances, our results are very suggestive. Increase in
clumping may be a general behaviour in SNe\,Ia. In the only case when extremely late
spectra have been obtained, SN\,2011fe, [\FeIII] emission is seen to disappear almost completely at epochs of 2 years and later, and is replaced by [\FeI] lines. Interestingly, the increased clumping implies roughly constant density within the clumps between d374 and d435, suggesting that clumps at that point start behaving like shrapnel. Clumping may be caused by the development of magnetic fields in the SN ejecta, and it may mark the beginning of the transition to a SNR.

{\bf Acknowledgements.} 
The authors thank TUBITAK, KFU, IKI, and  AST for partial support 
in using RTT150 (Russian-Turkish 1.5-m telescope in Antalya).
This work was partially funded by the Russian Government Program of
Competitive Growth of Kazan Federal University.

\end{document}